\documentclass[11pt,dvips]{article}
\usepackage{amsmath,amstext,amsbsy,amssymb}
\usepackage{bm}
\usepackage{graphicx}
\newcommand{\tr}{{\rm tr}}
\newcommand{\lsr}{\lambda_{\scriptscriptstyle{ R}}}
\newcommand{\dso}{\Delta_{\scriptscriptstyle{ SO}}}

\newcommand{\so}{{\scriptscriptstyle{ SO}}}

\newcommand{\ssR}{{\scriptscriptstyle{R}}}

\newcommand{\ssSH}{{\scriptscriptstyle{SH}}}
\newcommand{\ssSC}{{\scriptscriptstyle{SC}}}

\newcommand{\ssU}{{\scriptscriptstyle{U}}}

\newcommand{\ca}{\cal I}

\newcommand{\diag}{{\rm diag}}

\long\def\symbolfootnote[#1]#2{\begingroup%
\def\thefootnote{\fnsymbol{footnote}}\footnote[#1]{#2}\endgroup}

\begin{document}

\begin{center}

{\Large \bf The effective field theory of    $2+1$ dimensional topological insulator in the presence of  Rashba spin-orbit interaction  }

\vspace{2cm}

{\bf\"{O}mer F. Dayi},
 {\bf Mahmut Elbistan}\\
\vspace{5mm}

{\em {\it Physics Engineering Department, Faculty of Science and
Letters, Istanbul Technical University,\\
TR-34469, Maslak--Istanbul, Turkey} }
\vspace{5mm}

{dayi@itu.edu.tr , elbistan@itu.edu.tr}

\vspace{3cm}

{\bf Abstract}

\end{center}

 $2+1$ dimensional topological insulator described by the Kane-Mele model in the presence of Rashba spin-orbit interaction 
is considered. The effective action of the external fields coupled to  electromagnetic and spin degrees of freedom is accomplished within this model.
The Hamiltonian methods are adopted to  provide the coefficients appearing in the  action. 
It is demonstrated straightforwardly that the coefficients of the Chern-Simons terms are given by the first Chern number attained through the related non-Abelian Berry gauge field.
The effective theory which we obtain is in accord with the existence of the spin Hall phase where the value of  the spin Hall conductivity is very close to the quantized one.

\vspace{1cm}

\renewcommand{\theequation}{\thesection.\arabic{equation}}

\section{Introduction}

The Kane-Mele model of  monolayer graphene \cite{km} provides a formulation of the 2+1 dimensional  time reversal invariant topological insulator. In this  new topological phase of matter  which is also known as  the quantum spin Hall insulator, the bulk is insulating but there exist topologically protected gapless edge states. 
Charge carriers of graphene are effectively massless Dirac-like fermions at the Dirac points in the low energy and long wavelength regime. 
Kane and Mele explored properties of these electrons  in the presence of  intrinsic as well as Rashba spin-orbit interactions. 
In the Kane-Mele model when only the intrinsic spin-orbit coupling is considered,   two copies of the Haldane model \cite{hal} are combined to procure a quantized spin Hall conductivity. 
In \cite{km} it was argued that when both the intrinsic and Rashba coupling terms are present, although the spin Hall conductivity is not quantized it has a value which  slightly differs from the quantized one. Indeed, this is confirmed in \cite{ssth} by   numerical  methods. 
Because of  weak intrinsic spin-orbit coupling strength, the spin Hall phase in graphene is not  experimentally realizable \cite{minetal},\cite{yaoetal}. However, there are some recent proposals of synthesizing new materials which possess the  honeycomb structure of graphene with a large spin-orbit gap. For instance in \cite{taretal} a system of ultracold gas of potassium atoms was shown to provide an analogue of graphene.  There is also silicene which is a monolayer of Si atoms possessing the same lattice structure of graphene \cite{si}, whose low energy effective Hamiltonian is the same with the Kane-Mele model \cite{ljy}. Realization of  the molecular graphene \cite{mg} was another exiting achievement. In fact in \cite{ggh} it was argued that the molecular graphene can be used to realize the Kane-Mele quantum spin Hall model. 
In this construction presence of   the Rashba spin-orbit coupling term is essential, it cannot be switched off. 
Hence, it would be  very helpful to have a better understanding  of the
  main features of  Kane-Mele model in the presence of Rashba interaction.
 We  approach this problem from an effective action point of view.

Possessing 
effective field theory of the external fields coupled to charge and spin degrees of freedom is an efficient tool to reveal the general predictions of topological insulators \cite{qhz}.   Effective theories are  insensible to the internal structure of the inspected material, yet give its response  to the external fields. 
When only the intrinsic spin-orbit  coupling term is taken into account, the effective theory of   $2+1$ dimensional time reversal invariant topological insulator   is well established \cite{qhz},\cite{dey}. 
It is  a topological field theory where the coefficients  appearing in the action are related to the first Chern numbers of the constituting  
Dirac-like Hamiltonians. If one introduces only the  external electromagnetic gauge field, the effective Lagrange density is given by the Chern-Simons term whose coefficient  vanishes. This was expected because of the fact that the underlying model is time reversal invariant in contrast to the  Chern-Simons action which changes sign under this symmetry. However, one can also couple an external field  to the third component of spin which combines with the electromagnetic field to procure a non-vanishing effective action whose coefficient is the quantized spin Hall conductivity \cite{gs}. 
 In the presence  of  Rashba interaction  the third component of spin is not conserved, nevertheless
 one can still deal with the  spin Hall conductivity whose presence indicates the spin Hall phase. 
We would like to reveal  if the expected spin Hall conductivity can be obtained from the  effective 
action of the external electromagnetic and spin fields within the  Kane-Mele model in the presence of  Rashba interaction. 
This effective theory  
  was studied in terms of Lagrangian methods in \cite{srm} where the relevant coefficients derived at the first order in the Rashba coupling constant. 
We prefer to derive  the effective action of the external fields coupled to charge and spin degrees of freedom of the fully fledged Kane-Mele model
within  Hamiltonian methods  where the link between the coefficients taking part in the effective theory and topological Chern numbers can be discovered straightforwardly.

In the  next section we first  recall how one constructs the effective theory when  only  the intrinsic spin-orbit coupling term is present. Then we will discuss how to extend this method to obtain the effective action for the fully fledged Kane-Mele model. The main 
difficulty shows up in the  calculation of the coefficients which are defined in terms of Green functions. We would like to employ the Foldy-Wouthuysen transformation to obtain the one particle Green function for the Dirac-like Hamiltonian of  Kane-Mele model. 
In Section  \ref{coce} we explicitly construct the related Green functions and calculate explicitly the coefficients. 
Because of employing the Foldy-Wouthuysen transformation we  straightforwardly construct the non-Abelian Berry gauge field and  demonstrate   that the coefficients of Chern-Simons terms are given by the first Chern number. Calculation of the other coefficient
is cumbersome. 
Some details of this lengthy  calculation are reported in Appendix. The results which we obtained are in agreement with the existence of the spin Hall phase where the  spin Hall conductivity  possesses approximately the quantized value. In the last section we discuss
the results which we obtained as well as  a possible relation with  another approach.

\setcounter{equation}{0}
\section{Effective Field Theory}

Graphene has a honeycomb lattice structure based on two sublattices namely $A$ and $B$. At the two inequivalent Dirac points $K,\ K'$ of the Brillouin zone valence and conduction bands touch each other. Around these points in the low energy and long wavelength limit charge carriers effectively obey the free, massless Dirac-like Hamiltonian 
$$
H_0=\sigma_{x}\tau_{z}p_x +\sigma_yp_y,
$$
where we set the effective velocity of electrons equal to one, $v_F=1.$   
The Pauli spin matrices $\sigma_{x,y,z}$ act on the states of the sublattices $A,\ B,$ and  $\tau_z=\diag (1,-1)$  denotes the Dirac points
 $K,\ K'$.
We suppress the direct products between different spaces. In \cite{km}
Kane and Mele suggested to generate a mass gap by 
the intrinsic spin-orbit coupling term
$$
H_\so= \dso\sigma_{z}\tau_{z}s_{z}.
$$
They also considered  the Rashba spin-orbit interaction  term 
$$
H_\ssR=\lambda_{\ssR}(\sigma_{x}\tau_{z}s_{y}-\sigma_{y}s_{x}),
$$
where the constant parameter $\lambda_\ssR $ is  experimentally tunable.  
The Pauli spin matrices $s_{x,y,z}$  correspond to the spin degrees of freedom of electrons.
Hence, the Hamiltonian of the Kane-Mele model including intrinsic as well as Rashba spin-orbit interactions  is
\begin{equation}
\label{H}
H    =   H_{0}+H_\so+H_\ssR.
\end{equation}

For $\lambda_{\ssR}=0$ the third component of spin, $s_z,$  which can be labeled by $\uparrow\downarrow$   is conserved. Thus, spin current can directly be defined by
 $j^{spin}=j^\uparrow -j^\downarrow .$ It leads to 
 the quantized spin Hall conductivity  $\sigma_{\ssSH}=1/2\pi ,$  in the $e=1,\ \hbar=1$ units. 
This spin current can also be derived from the action, $\mu ,\nu ,\rho =0,1,2,$
\begin{equation}
\label{EfL0}
S_{s}=\frac{1}{2\pi}\int d^3x \epsilon^{\mu\nu\rho}\Omega_\mu\partial_\nu A_\rho ,
\end{equation}
where  $A_\mu$ and $\Omega_\mu$ are the external fields  associated with the electromagnetic and the spin \mbox{currents \cite{gs}.  }
As we will discuss above, (\ref{EfL0})  results  as the effective action obtained by integrating out the fermionic fields in the  
path integral of the 
field theory described by the following Lagrangian density
of the Kane-Mele model for $\lambda_{\ssR}=0,$
\begin{equation}
{\cal L}_0 =
\bar{\psi} \left[ \gamma^\mu \left( i\partial_\mu + A_\mu +\frac{S_z}{2} \Omega_\mu \right)- \dso \right] \psi .
\label{lagL}
\end{equation}
Here $\gamma^0=\sigma_{z}\tau_{z}s_{z},\ \gamma^1=i\sigma_{y}s_{z},\  \gamma^2=-i\sigma_{x}\tau_{z}s_{z},$ and 
$
S_z={\rm diag }\left(s_z,s_z,s_z,s_z\right).
$

We would like to extend this stratagem for deriving the spin Hall conductivity  to the fully fledged Kane-Mele model given by (\ref{H}).
Although when  $\lambda_{\ssR}$ is nonvanishing the third component of spin does not commute with the Hamiltonian
 (\ref{H}), 
so that the current  $j^s_\mu=\bar{\psi} \gamma^\mu S_z \psi /2,$ is not conserved,
one can still define a  spin current  which is conserved in the low energy limit (for a similar approach see \cite{mnz}) and calculate the spin Hall conductivity. Indeed, Kane and Mele argued that when $\dso >\lambda_{\ssR}$  the Hamiltonian (\ref{H}) yields the spin Hall conductivity which slightly differs from the quantized value
 $1/2\pi .$  This is confirmed in \cite{ssth} by studying the model numerically.  We  approach 
this problem from another point of view. We would like to derive the effective field theory of external fields   $A_\mu,\Omega_\mu, $
considering the Kane-Mele model   Lagrange density in the presence of Rashba spin-orbit interaction:
\begin{equation}
 {\cal L} \left(\psi,\bar{\psi}, A , \Omega\right) =
\bar{\psi} \left[ \gamma^\mu \left( i\partial_\mu + A_\mu +\frac{S_z}{2} \Omega_\mu \right)- \dso -\lsr(\sigma_{y}s_{x}-\sigma_{x}\tau_zs_{y})\right] \psi .
\label{lag}
\end{equation}
In the partition function,
$$
Z=\int D\psi D\bar{\psi} DA_{\mu} D\Omega_{\mu}e^{i\int d^3x{\cal L}} ,
$$
we may integrate out $\psi$ and $\bar{\psi} $ to acquire the effective theory of 
the external fields $A_\mu , \Omega_\mu$:
$$
\int D\psi D\bar{\psi} DA_{\mu} D\Omega_{\mu}e^{iS}=\int DA_{\mu} D\Omega_{\mu}e^{iS_{eff}} .
$$
$S_{eff}$ is defined as
\begin{equation}
\label{seffdet}
S_{eff}[A,\Omega]=-i\ln\det \left[i\gamma^{\mu}(\partial_{\mu}-iA_{\mu}-i\frac{S_z}{2}\Omega_{\mu})-\dso-\lsr(\sigma_{y}s_{x}-\sigma_{x}\tau_zs_{y})\right] .
\end{equation}
We are interested  only in  the following terms which (\ref{seffdet}) evokes in the low energy limit,
\begin{equation}
\label{Ef}
S_{eff}=C\int d^3x \epsilon^{\mu\nu\rho}A_\mu\partial_\nu A_\rho +C_s\int d^3x \epsilon^{\mu\nu\rho}\Omega_\mu\partial_\nu A_\rho
+C_\Omega \int d^3x \epsilon^{\mu\nu\rho}\Omega_\mu\partial_\nu \Omega_\rho.
\end{equation}
This action yields the spin current 
$$
j^{\mu}_{spin}=\frac{\delta S_{eff}}{\delta \Omega_{\mu}}.
$$ 
It is worth mentioning that 
the spin current obtained from the action (\ref{Ef}) is conserved 
\mbox{$\partial_\mu  j^{\mu}_{spin}=0,$}
though  the third component of spin $S_z$ does not commute with the Hamiltonian (\ref{H}). It is a consequence of dealing with the low energy limit where the higher order gradient terms  in the expansion of  (\ref{seffdet}) are ignored. 

In the weak field approximation the coefficients in (\ref{Ef}) are 
given in terms  of the fermion propagator $G(p)$ and its inverse $G^{-1}(p)$  by \cite{gjk} 
\begin{eqnarray}
\label{ca}
C & = & -\frac{1}{12}\epsilon^{\mu\nu\rho}\int \frac{d^3p}{(2\pi)^3}\tr \left\{\left[G(p)\partial_{\mu}G^{-1}(p)\right]\left[G(p)\partial_{\nu}G^{-1}(p)\right]\left[G(p)\partial_{\rho}G^{-1}(p)\right]\right\} ,\\
\label{cs}
C_{s}& =& -\frac{1}{12}\epsilon^{\mu\nu\rho}\int \frac{d^3p}{(2\pi)^3}\tr\left\{S_z\left[G(p)\partial_{\mu}G^{-1}(p)\right]\left[G(p)\partial_{\nu}G^{-1}(p)\right]\left[G(p)\partial_{\rho}G^{-1}(p)\right]\right\} ,\\
\label{cO}
C_\Omega & = &-\frac{1}{12}\epsilon^{\mu\nu\rho}\int \frac{d^3p}{(2\pi)^3}\tr\left\{S_z\left[G(p)\partial_{\mu}G^{-1}(p)\right]S_z\left[G(p)\partial_{\nu}G^{-1}(p)\right]\left[G(p)\partial_{\rho}G^{-1}(p)\right]\right\} ,
\end{eqnarray}
where $\partial_\mu\equiv \partial/ \partial p^\mu .$ For $\lambda_{\ssR}=0,$  we can  write 
$H_0+H_\so=\diag (H^{\uparrow +}, H^{\uparrow -},H^{\downarrow +},H^{\downarrow -})$  
in terms of $2\times 2 $ matrices, where $\pm$ labels the Dirac points $K,\ K^\prime.$  One can show that the
coefficients can be expressed in terms of the related first Chern numbers  (\cite{qhz}, \cite{dey} and the references therein). In fact the coefficients 
of the Chern-Simons terms are given by 
$$
C (\lambda_{\ssR}=0)=C_\Omega(\lambda_{\ssR}=0) =(N_1^{\uparrow +}+ N_1^{\uparrow -}+N_1^{\downarrow +}+N_1^{\downarrow -})/4\pi.
$$
The related first Chern numbers were obtained to be  $N_1^{\uparrow \pm }=1/2,\ N_1^{\downarrow \pm }=-1/2.$ Thereby one observes that $C (\lambda_{\ssR}=0)=C_\Omega(\lambda_{\ssR}=0) =0.$ Vanishing of these coefficients was expected due to the fact that Kane-Mele model is time reversal invariant but the Chern-Simons terms lack this symmetry.
However, the other coefficient is given by 
$$
C_s(\lambda_{\ssR}=0) =(N_1^{\uparrow +}+ N_1^{\uparrow -}-N_1^{\downarrow +}-N_1^{\downarrow -})/4\pi=1/2\pi .
$$
Therefore, (\ref{EfL0}) occurs to be the effective action of the theory described by  (\ref{lagL}).

In the presence of  Rashba interaction, i.e. $\lambda_{\ssR}\neq 0,$ the coefficients (\ref{ca})-(\ref{cO}) were  constructed within Lagrangian methods in \cite{srm} up to the first order terms in $\lambda_{\ssR}/ \dso.$ Moreover,
 in \cite{cgv} an effective theory for a  model which is similar to the Kane-Mele model\footnote{ In \cite{cgv} it was claimed that the model considered  is equivalent to the Kane-Mele model (\ref{H}), because of  only  employing  another representation of the gamma matrices given by $\gamma^\mu =\sigma_z\gamma_{KM}^\mu ,$ where  $\gamma_{KM}^\mu$ are the gamma matrices of the Kane-Mele model. However, they adopted the definition $\gamma^0=\sigma_z$ which  leads to the erroneous result $\gamma_{KM}^0=1.$ In fact multiplying  the Pauli matrices by $\sigma_z$ yields an equivalent set of matrices if one  takes into account also the identity matrix $1_\sigma,$ i.e. multiplying the set of matrices $(\sigma_ x,\sigma_y ,\sigma_z ,1_\sigma)$ by $\sigma_z$ one gets the equivalent set of matrices $(i\sigma_ y,-i\sigma_x ,1_\sigma ,\sigma_z ).$}
was  constructed. The following section is devoted to the explicit calculations of these coefficients.

\setcounter{equation}{0}
\section{Calculation of the Coefficients \label{coce}}

To attain the one particle Green function of  free Dirac field $G(p),$ we would like to employ the Hamiltonian methods.
We choose to order the direct products such that the explicit form of the Kane-Mele model Hamiltonian (\ref{H})  becomes
\begin{equation}
\label{HA}
H=\left( 
\begin{array}{cccccccc}
\dso & 0 & 0 & 0 & p_x-ip_y & 0 & 0 & 0 \\
0 & -\dso & 0 & 0 & 2i\lsr & p_x-ip_y & 0 & 0 \\
0 & 0 & -\dso & 0 & 0 & 0 & -p_x-ip_y & 2i\lsr \\
0 & 0 & 0 & \dso & 0 & 0 & 0 & -p_x-ip_y \\
p_x+ip_y & -2i\lsr & 0 & 0 & -\dso & 0 & 0 &0 \\
0 & p_x+ip_y & 0 & 0 & 0 & \dso & 0 & 0 \\
0 & 0 & -p_x+ip_y & 0 & 0 & 0 & \dso & 0 \\
0 & 0 & -2i\lsr & -p_x+ip_y & 0 & 0 & 0 & -\dso
\end{array}
\right) .
\end{equation}
In terms of $p^2=p_x^2+p_y^2,$ the eigenvalues of (\ref{HA}) are calculated  to be 
\begin{equation}
\label{eig}
\begin{array}{rl}
E_1=E_2=&\lsr+\sqrt{(\dso-\lsr)^2+p^2} , \\
E_3=E_4=&-\lsr+\sqrt{(\dso+\lsr)^2+{p}^2}, \\
E_5=E_6=&\lsr-\sqrt{(\dso-\lsr)^2+{p}^2},\\
E_7=E_8=&-\lsr-\sqrt{(\dso+\lsr)^2+{p}^2}.
\end{array}
\end{equation} 
$G(p)$ can be acquired by means of   the  Foldy-Wouthuysen unitary transformation $U$ which is defined to satisfy
\begin{equation}
\label{diag}
UHU^\dagger =\diag (E_1,\cdots ,E_8)\equiv\sum_{M=1}^8{E_{M} I^M}.
\end{equation} 
Here $I^M$ is the matrix whose elements vanish other than  $(I^M)_{MM}=1.$  
The eigenfunctions  corresponding to the energy eigenvalues (\ref{eig})  can be employed to establish the unitary matrix $U$   which diagonalizes the Hamiltonian (\ref{HA}), as follows,
\begin{equation}
\label{UU}
U=\left(
\begin{array}{cccccccc}
0 & 0 & -iF_1 & \frac{(p_x-ip_y)F_1}{\dso-E_1} & 0 & 0 & \frac{-i(p_x+ip_y)F_1}{\dso-E_1} & F_1 \\
\frac {i(p_x+ip_y)F_2}{p_x-ip_y} & \frac{-(\dso-E_2)F_2}{p_x-ip_y} & 0 & 0 & \frac{-i(\dso-E_2)F_2}{p_x-ip_y} & F_2 & 0 & 0 \\
0 & 0 & iF_3 & \frac{(p_x-ip_y)F_3}{\dso-E_3} & 0 & 0 & \frac{i(p_x+ip_y)F_3}{\dso-E_3} & F_3 \\
\frac {-i(p_x+ip_y)F_4}{p_x-ip_y} & \frac{-(\dso-E_4)F_4}{p_x-ip_y} & 0 & 0 & \frac{i(\dso-E_4)F_4}{p_x-ip_y} & F_2 & 0 & 0 \\
0 & 0 & -iF_5 & \frac{(p_x-ip_y)F_5}{\dso-E_5} & 0 & 0 & \frac{-i(p_x+ip_y)F_5}{\dso-E_5} & F_5 \\
\frac {i(p_x+ip_y)F_6}{p_x-ip_y} & \frac{-(\dso-E_6)F_6}{p_x-ip_y} & 0 & 0 & \frac{-i(\dso-E_6)F_6}{p_x-ip_y} & F_6 & 0 & 0 \\
0 & 0 & iF_7 & \frac{(p_x-ip_y)F_7}{\dso-E_7} & 0 & 0 & \frac{i(p_x+ip_y)F_7}{\dso-E_7} & F_7 \\
\frac {-i(p_x+ip_y)F_8}{p_x-ip_y} & \frac{-(\dso-E_8)F_8}{p_x-ip_y} & 0 & 0 & \frac{i(\dso-E_8)F_8}{p_x-ip_y} & F_8 & 0 & 0 
\end{array}
\right).
\end{equation}
The normalization factors are given by
\begin{eqnarray}
\label{f}
F_{2m-1}=\sqrt{\frac{(\dso-E_m)^2}{2((\dso-E_m)^2+p^2)}},\ F_{2m}=\sqrt{\frac{p^2}{2((\dso-E_m)^2+p^2)}};\ \ 
m=1,2,3,4.
\end{eqnarray}
Observe that they satisfy
\begin{equation}
\label{f2}
F_{2m-1}^2+F_{2m}^2=\frac{1}{2}.
\end{equation}
Inverting the unitary transformation (\ref{diag}), we can retrieve  the Hamiltonian (\ref{HA}) in the form
$$
H=\sum_{M=1}^8{E_{M} P^M},
$$
where we introduced
$
P^M=U^\dagger I^M U. 
$
 Obviously $P^M$ are projection operators: 
$$
\sum_{M=1}^8 P^M=1,\ \ P^MP^N=\delta^{MN}P^N.
$$
Now, one can construct the Green function $G(p)$ and its inverse $G^{-1}(p)$ as
\begin{equation}
\label{gin}
G(p)=\sum_{M=1}^8\frac{P^M}{w-E_M},\ G^{-1}(p)=w-\sum_{M=1}^8 E_MP^M ,
\end{equation} 
where $p^\mu =(w, p_a);\ a=1,2.$
Note that the derivatives of the inverse  Green function $G^{-1}(p)$ obey 
\begin{equation}
\label{dervi}
\frac{\partial G^{-1}(p)}{\partial w}=1,\ \frac{\partial G^{-1}(p)}{\partial p_a}=-\sum_{M=1}^8\left(\frac{p_a}{E_M}P^M+E_M\partial_aP^M\right) .
\end{equation} 
To proceed we would like to perform the $w$ integrations in (\ref{ca})-(\ref{cO}). This requires that  the energies $E_M;\ M=1,\cdots, 8,$ are
arranged to be definitely positive or negative. We restrict the values of coupling constant  to $\dso>2\lsr,$  so that we can divide the spectrum as $E_{\alpha},\ \alpha=1,2,3,4, $ which are positive and $E_i,\ i=5,6,7,8,$ which are negative. Notice that in the limit $\lsr\to 0$, the eigenvalues $E_{1,2}$ and $E_{3,4}$ (similarly $E_{5,6}$ and $E_{7,8}$)  approach each other. In the following  we adopt the conventions: 
$\alpha,\beta, \gamma =1,...,4; $ $i,j,k= 5,...,8;$   $M, N=1,...,8;$ and $a, b=1,2.$ 

\subsection{Calculation of $C$ and $C_\Omega$}
The winding number  (\ref{ca}) can be written  as
\begin{equation}
\label{winu}
C=-\frac{1}{4}\varepsilon^{ab}\int \frac{d^2pdw}{(2\pi)^3}\tr\left\{G^2(p)\partial_{a}G^{-1}(p)G(p)\partial_{b}G^{-1}(p)\right\} .
\end{equation}
The repeating $a$ and $b$ indices are summed over.
Employing the Green function (\ref{gin}) and the derivatives of its inverse given in (\ref{dervi})  one can show  that the terms explicitly linear and quadratic  in $p_a$ vanish directly and the  rest  after performing the $w$ integration leads to
$$
C=\frac{-i}{16\pi^2}\varepsilon^{ab}\int d^2p \tr \left\{\sum_{\alpha,\beta,i}\frac{E_{\beta}P^i}{E_{\alpha}-E_i}\left(\partial_aP^{\alpha}\partial_bP^{\beta}-\partial_bP^{\beta}\partial_aP^{\alpha}\right)
+ \sum_{\alpha,i,j}\frac{E_jP^{\alpha}}{E_{\alpha}-E_i}\Big(\partial_aP^i\partial_bP^j-\partial_bP^j\partial_aP^i\Big)  \right\} .
$$
The terms other than  $\alpha=\beta$ and $i=j$ do not contribute, hence one gets
\begin {equation}
\label{ccc}
C=-\frac{i}{8 \pi^2}\varepsilon^{ab}\sum_{\alpha, i} \int d^2p \tr\left\{\frac{E_{\alpha}}{E_{\alpha}-E_{i}}P^i\partial_a P^{\alpha}\partial_b P^{\alpha}+\frac{E_i}{E_{\alpha}-E_i}P^{\alpha}\partial_a P^i\partial_b P^i\right\}.
\end{equation}
Recalling that  $P^M=U^\dagger I^M U,$ (\ref{ccc}) can be expressed in the form
\begin{equation}
C=\frac{i}{8 \pi^2}\varepsilon^{ab}\sum_{\alpha,i} \int d^2p \tr \left\{I^\alpha A^\ssU_a I^i A^\ssU_b\right\},
\label{afo}
\end{equation}
where we introduced  $A^\ssU_a=i U \partial_a U^\dagger .$ 
Because of being
 a pure gauge field its curvature   identically vanishes. However, we can construct the 
 Berry gauge field through the  projection of the  $A^\ssU$ to the positive energy states \cite{bm} by
$$
A^B_a=i\sum_{\alpha, \beta} I^{\alpha}U \partial_a U^\dagger I^{\beta},
$$  
 whose  field strength 
$$
F^B_{ab}=\partial_a A^{B}_b-\partial_b A^{B}_a-i[A^{B}_a, A^{B}_a],
$$
does not vanish in general.
The first Chern number is defined in terms of  the Berry curvature as
\begin{equation}
\label{N1}
N_1=\frac{1}{4\pi}\int d^2p \varepsilon^{ab} \tr   F^B_{ab}.
\end{equation}
The  topological numbers (\ref{winu}) and (\ref{N1})
are connected to each other by
$$
C=\frac{N_1}{4\pi}.
$$ 
This relation can be accomplished by observing that, due to  the identity 
$
\sum_i I^i=1-\sum_\alpha I^\alpha ,
$
one can  express (\ref{afo}) as 
\begin{equation}
\label{ca2}
C=-\frac{i}{8 \pi^2}\varepsilon^{ab}\sum_{\alpha}\int d^2p \tr \left\{I^{\alpha}U \partial_a U^\dagger U \partial_b U^\dagger I^{\alpha}\right\} 
+\frac{i}{8 \pi^2}\varepsilon^{ab}\sum_{\alpha,\beta} \int d^2p \tr \left\{I^{\alpha}U \partial_a U^\dagger I^{\beta}U \partial_b U^\dagger I^{\alpha}\right\} .
\end{equation}

Having attained the relation  between the Chern number (\ref{N1}) and the winding number (\ref{winu}),  the next step is to calculate  
the coefficient $C$ explicitly. In  a straightforward manner  (\ref{ca2}) can be written as
\begin{equation}
\label{ccal}
C=\frac{i}{8\pi^2}\varepsilon^{ab}\int \tr\left\{\sum_{\alpha}I^{\alpha}\partial_aU\partial_bU^{\dagger}I^{\alpha}-\sum_{\alpha,\beta}I^{\alpha}\partial_aUP^{\beta}\partial_bU^{\dagger}I^{\alpha} \right\}.
\end{equation}
After performing the trace operation and making use of (\ref{UU}), the first term  vanishes:
$$
\partial_aU_{21}\partial_bU^{\star}_{21}+2\partial_aU_{22}\partial_bU^{\star}_{22}+\partial_aU_{41}\partial_bU^{\star}_{41}+2\partial_aU_{42}\partial_bU^{\star}_{42}=0. 
$$
 On the other hand, the second term in (\ref{ccal})  yields
\begin{eqnarray}
\varepsilon^{ab}\big\{(P^2_{11}+P^4_{11})\left(\partial_aU_{21}\partial_bU^{\star}_{21}+\partial_aU_{41}\partial_bU^{\star}_{41}\right) \nonumber\\
+4i{\rm Im}\left[P^2_{12}\left(\partial_aU_{21}\partial_bU^{\star}_{22}+\partial_aU_{25}\partial_bU^{\star}_{26}\right)
+P^4_{12}\left(\partial_aU_{41}\partial_bU^{\star}_{42}+\partial_aU_{45}\partial_bU^{\star}_{46}\right)\right]  \label{ca3}\\
+2i{\rm Im}[(P^2_{16}+P^4_{16})\left(\partial_aU_{21}\partial_bU^{\star}_{26}+\partial_aU_{41}\partial_bU^{\star}_{46}\right)]
+4P^2_{22}\partial_aU_{22}\partial_bU^{\star}_{22}+4P^4_{22}\partial_aU_{42}\partial_bU^{\star}_{42}) \big\}. \nonumber
\end{eqnarray}
In terms of the polar coordinates
\begin {equation}
\label{pol}
p=\sqrt{p_x^2+p_y^2},\ \theta=\arctan\frac{p_y}{p_x},
\end{equation}
one can demonstrate that (\ref{ca3}) vanishes as 
$$
\frac{4i}{p}(F_2^2\partial_pF_2^2+F_4^2\partial_pF_4^2)-\frac{8i}{p}(F_2^3\partial_pF_2+F_4^3\partial_pF_4)]=0.
$$

In these calculations we have utilized the explicit forms of  Green functions obtained from the Kane-Mele model. However, the properties
of Green functions which led us to conclude that the coefficient $C$ vanishes, are extendable to any Dirac-like theory whose energy spectrum possesses particle-hole (antiparticle) symmetry.

One of the benefits of using Hamiltonian methods is the fact that
the coefficients corresponding to the subspaces labeled by $\tau_z=\pm 1$ and $s_z= \uparrow \downarrow $
can be calculated  explicitly. In fact they yield the Chern numbers
$$
N_1^{\uparrow \pm }=1/2,\ N_1^{\downarrow \pm }=-1/2.
$$

The coefficient $C_\Omega$ (\ref{cO}),  can  be demonstrated to be equal to $C$ (\ref{ca}): Observe that
$$
S_zH(\lambda_\ssR)S_z= H(-\lambda_\ssR).
$$
This interchanges the positive and negative energy eigenvalues within themselves: $E_{1} \leftrightarrow E_{3}$ and $E_{5} \leftrightarrow E_{7}.$
Thereby $S_zG^{-1}S_z$ and $S_zGS_z$ are captured from $G^{-1}$ and $G$ by this exchange of eigenvalues. However,
the initial $w$ integrals which led to (\ref{ccc}) are not altered under the exchange $E_{1} \leftrightarrow E_{3}$ and $E_{5} \leftrightarrow E_{7}.$ Therefore one concludes that  $C_\Omega =C=0.$ 

Vanishing of the coefficients $C_\Omega$ and $C,$ was expected
 because of the fact that under the time reversal symmetry
 Kane-Mele model is  invariant  but  Chern-Simons action acquires an overall minus  sign.

\subsection{Calculation of $C_s$}

Substituting the first derivative of the inverse Green function with respect to $w$ by (\ref{dervi}), the coefficient $C_s,$  (\ref{cs}),
can be separated into three parts as
\begin{eqnarray}
\label{cs3}
C_{s}&=&-\frac{1}{12}\varepsilon^{ab}\int \frac{d^2pdw}{(2\pi)^3}\tr\Big\{S_zG^2(p)\partial_aG^{-1}(p)G(p)\partial_bG^{-1}(p)-S_zG(p)\partial_aG^{-1}(p)G^2(p)\partial_bG^{-1}(p) \nonumber \\
&&+S_zG(p)\partial_aG^{-1}(p)G(p)\partial_bG^{-1}(p)G(p)\Big\}  
\equiv C_{s}^{(1)}+C_{s}^{(2)}+C_{s}^{(3)} . 
\end{eqnarray}
Insertion of  $\partial_aG^{-1}(p)$ given by (\ref{dervi}) into  (\ref{cs3})  yields terms which are explicitly linear and quadratic in $p_a.$ 
Apparently, the terms quadratic in $p_a$ vanish. After some calculations one can show  on general grounds that the terms linear in $p_a$ also yield a vanishing contribution. To deal with the remaining terms,
 we first would like to perform the $w$ integration. For this aim,  we write the first and the second constituents of (\ref{cs3}) as
\begin{eqnarray*}
C_{s}^{(1)} & = & -\frac{1}{12}\varepsilon^{ab}\sum_{\alpha,i,M}\int\frac{d^2pdw}{(2\pi)^3}\tr\left\{S_z\left(\frac{E_iP^{\alpha}\partial_aP^i}{(w-E_{\alpha})^2(w-E_i)}-\frac{E_iP^i\partial_aP^{\alpha}}{(w-E_i)^2(w-E_{\alpha})}\right)E_M\partial_bP^M \right\},\\
C_{s}^{(2)}  & = & \frac{1}{12}\varepsilon^{ab}\sum_{\alpha,i,M}\int\frac{d^2pdw}{(2\pi)^3}\tr\left\{S_z\left(\frac{E_iP^{\alpha}\partial_aP^i}{(w-E_{\alpha})(w-E_i)^2}-\frac{E_iP^i\partial_aP^{\alpha}}{(w-E_i)(w-E_{\alpha})^2}\right)E_M\partial_bP^M \right\}.
\end{eqnarray*}
Now we can integrate over  $w$
and find that they acquire the same form:
$$
C_{s}^{(1)}=C_{s}^{(2)}=-\frac{i}{48\pi^2}\varepsilon^{ab}\sum_{\alpha,i,M}\int d^2p \tr \Big\{S_z\frac{P^{\alpha}\partial_aP^i+P^i\partial_aP^{\alpha}}{E_{\alpha}-E_i}E_M\partial_bP^M\Big\} .
$$
They can  be expressed as
\begin {equation}
\label{c1+c2}
C_{s}^{(1)}=C_{s}^{(2)}
=-\frac{i}{48\pi^2}\sum_{\alpha,i,M}\int d^2p \tr\left\{\frac{(E_i-E_M)P^MS_zP^{\alpha}{\ca}^i
+(E_{\alpha}-E_M)P^MS_zP^i{\ca}^\alpha}{E_{\alpha}-E_i}\right\},
\end{equation}  
where we defined  ${\ca}^M=\varepsilon^{ab}\partial_aU^{\dagger}I^M\partial_bU.$
However, due to the fact that $[S_z,P^M]\neq 0,$ the third constituent of (\ref{cs3}) yields  
$$
C_{s}^{(3)}=-\frac{\varepsilon^{ab}}{12}\sum_{L,M\neq N}\int\frac{d^2p d\omega}{(2\pi)^3}
\tr\left\{\frac{(E_N-E_M)P^LS_zP^M}{(\omega-E_L)(\omega-E_M)(\omega-E_N)}
\left( E_L\partial_aP^N\partial_bP^L +E_N\partial_aP^N\partial_bP^N
\right)
\right\} .
$$
 By performing   the $w$ integration we get
\begin{eqnarray}
C_{s}^{(3)} &=& -\frac{i}{48\pi^2}\int d^2p\tr\Big\{
P^{57}\left({\ca}^1+{\ca}^3\right)+P^{68}\left({\ca}^2+{\ca}^4\right)-P^{13}\left({\ca}^5+{\ca}^7\right)-P^{24}\left({\ca}^6+{\ca}^8\right)\nonumber\\
&& +P^{35}\left(\frac{E_1-E_3}{E_5-E_3}{\ca}^1-\frac{E_7-E_5}{E_3-E_5}{\ca}^7\right)+P^{46}\left(\frac{E_2-E_4}{E_6-E_4}{\ca}^2-\frac{E_8-E_6}{E_4-E_6}{\ca}^8\right)\label{c33}\\
&& +P^{17}\left(\frac{E_3-E_1}{E_7-E_1}{\ca}^3-\frac{E_5-E_7}{E_1-E_7}{\ca}^5\right)+P^{28}\left(\frac{E_4-E_2}{E_8-E_2}{\ca}^4-\frac{E_6-E_8}{E_2-E_8}{\ca}^6\right)\
\Big\},\nonumber
\end{eqnarray}
where we introduced  $P^{MN}\equiv P^MS_zP^N+P^NS_zP^M.$ 
Combining (\ref{c33}) with (\ref{c1+c2})  we obtain $C_{s}$ as it is presented in Appendix. One finally   obtains
\begin{eqnarray}
C_{s}=&\frac{-2}{3\pi}\int dp\Big\{F_6^2F_8^2\left(\left [-3E_1E_3 -3E_3^2+3E_1^2+3(E_5 + E_7)^2+E_5^2+ E_7^2+E_5E_7\right]\frac{\partial_pF_1^2}{(\dso-E_1)^2}+1\leftrightarrow 3\right) &\nonumber\\
&-F_2^2F_4^2\left(\left[-3E_5E_7+3E_5^2-3E_7^2+3(E_1+E_3)^2+E_1^2+E_3^2+E_1E_3\right] \frac{\partial_pF_5^2}{(\dso-E_5)^2}+5\leftrightarrow 7\right) &\nonumber\\
&-\frac{F_4^2F_6^2}{E_5-E_3} \left((E_3-E_1)^3\frac{\partial_pF_1^2}{(\dso -E_1)^2 }- (E_7-E_5)^3\frac{\partial_pF_7^2}{(\dso -E_7)^2 }\right) &\nonumber\\
&+\frac{ F_2^2F_8^2}{E_7-E_1} \left((E_3-E_1)^3\frac{\partial_pF_3^2}{(\dso -E_3)^2}-(E_7-E_5)^3\frac{\partial_pF_5^2}{(\dso -E_5)^2} \right)\Big\}, &\label{csint} 
\end{eqnarray}
where $M\leftrightarrow N$ denotes  the term which arises from the former entry by interchanging $M$ and $N.$
Although we could not analytically solve the integral in (\ref{csint}), numerical calculations are in accord with 
the result
\begin{equation}
\label{csfinal}
C_{s} =R(\infty)-R(0),
\end{equation}
where $R(p)$ is deduced to be
\begin{eqnarray}
\label{infin}
R(p)&=&\frac{1}{6\pi}\Big[\frac{\dso-\lsr}{\sqrt{(\dso-\lsr)^2+p^2}}+\frac{\dso+\lsr}{\sqrt{(\dso+\lsr)^2+p^2}}\nonumber \\ 
&-&\frac{\dso}{2\lsr}\tanh^{-1}{\sqrt{1+\frac{p^2}{(\dso-\lsr)^2}}+\frac{\dso}{2\lsr}\tanh^{-1}{\sqrt{1+\frac{p^2}{(\dso+\lsr)^2}}}}\ \Big] .
\end{eqnarray}  
In the limit   $p\to \infty$ (\ref{infin}) vanishes, but its
$p \to 0$ limit depends on the ratio of the coupling constants $\dso$ and $\lsr$ as:
$$
\lim_{p\to 0} R(p)=\frac{1}{3\pi}\Big[1+\frac{1}{4}\frac{\dso}{\lsr}\ln(\frac{\dso+\lsr}{\dso-\lsr})\Big]
$$ 
 For diverse values of the coupling constants satisfying $\dso>2\lsr,$  $C_s$ occurs to be in the range between $0.506/ \pi$ 
and $0.500/ \pi .$ For example we find  $C_s=0.506/ \pi$  for $\frac{\dso}{\lambda_\ssR}=3$ and   $C_s=0.502/ \pi$ for $\frac{\dso}{\lambda_\ssR}=5.$ As we plotted in Figure 1, for  $\frac{\dso}{\lambda_\ssR}\geq 5,$  (\ref{csfinal}) has values closer to   $C_s=1/2\pi$ which is the   
exact result  when $\dso\gg\lsr.$

%%%%%%%%%%%%%%%%%%%%%%%%%%%%%%%%%%%%%%%%%%%%%
\begin{figure}
\begin{center}
\label{csF}
\includegraphics{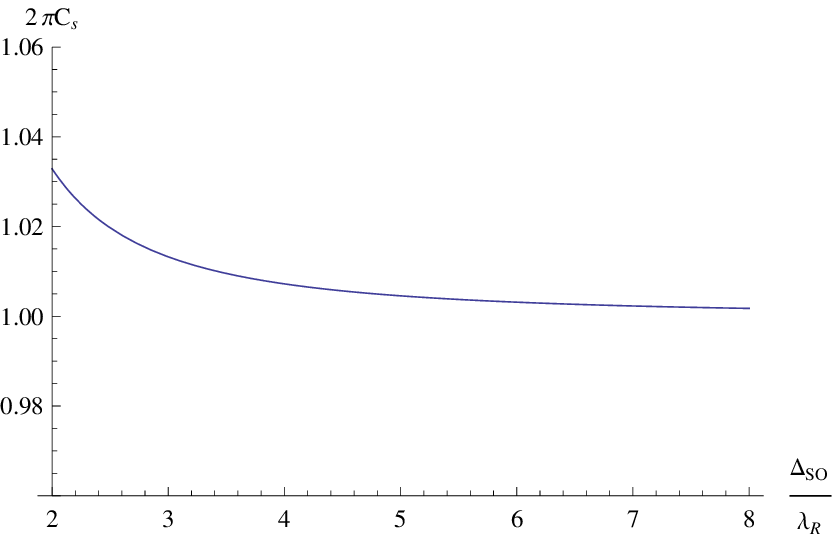}
\end{center}
\caption{The  coefficient $C_s$ with respect to the ratio $\dso/\lsr.$}
\end{figure}
%%%%%%%%%%%%%%%%%%%%%%%%%%%%%%%%%%%%%%%%%%%%%%%%

\setcounter{equation}{0}
\section{Discussions}

Response of the system (\ref{lag}) to the field $\Omega_\mu ,$ 
 provided  by the effective action (\ref{Ef}) is
$$
 j^{\mu}_{spin}=C_s\epsilon^{\mu\nu\rho}\partial_\nu A_\rho,
$$
where $C_s\approx e/2\pi. $ 
We do not deal with the quantum corrections to the  $BF$-type action (\ref{Ef}). Otherwise,  in the expansion of  
(\ref{seffdet}) we need to consider the higher gradient terms which we
ignored  in the low energy limit. The spatial component of  spin current
 can be interpreted in terms of the spin Hall conductivity $\sigma_\ssSH$ and the electric field $E_a=\partial_a A_0-\partial_0A_a$ as
$$
j_{spin}^a=\sigma_\ssSH \epsilon_{ab}E_a.
$$
Thus, we can conclude that for the Kane-Mele model  in the presence of Rashba interaction (\ref{lag})  it has approximately the quantized value
\begin{equation}
\sigma_\ssSH =C_s \approx \frac{1}{2\pi},
\label{br}
\end{equation}
for $\dso>2\lsr$. 

 Within  the Kane-Mele model in the presence of  Rashba 
interaction, (\ref{lag}), a  realization of the $2+1$ dimensional  spin Hall phase  in a certain range of values of the interaction parameters,  $\dso,\ \lsr,$ was suggested in \cite{ggh}. 
 In this molecular graphene construction   to achieve quantum spin Hall phase a
suitable  set of values  was given by $\dso=0.145 eV,$ and $ \lsr=0.04 eV.$   In fact, adopting
these values our numerical calculation  produces the result  $\sigma_\ssSH \approx 1/2\pi .$ 

Moreover, response of the Kane-Mele model  (\ref{lag})  to the external  electromagnetic gauge field $A_\mu$ 
can be derived from the effective action (\ref{Ef}) as
$$
j^{\mu}_{charge}= \frac{\delta S_{eff}}{\delta A_{\mu}}.
$$
We have shown that the coefficient of the Chern-Simons term, $C,$ vanishes, so that 
the charge current furnished by the effective action is
\begin{equation}
j^{\mu}_{charge}\approx \frac{1}{2\pi}\epsilon^{\mu\nu\rho}\partial_\nu \Omega_\rho .
\label{cjm}
\end{equation}
In \cite{qhz} it was demonstrated that (\ref{cjm}) is the fundamental response equation for the quantum spin Hall  effect.

To consider the spin Hall phase a topological invariant 
 spin Chern number was introduced   in \cite{pro}. The definition of \cite{pro} relies on the fact that one can 
project eigenstates of a gapped Hamiltonian to  up and  down sectors of the spin operator $S_z,$ even if it does not commute with the related Hamiltonian.  In \cite{yetal}   this definition was adopted to calculate the spin Chern numbers $N^{\ssSC}_{\pm \uparrow  \downarrow }$  for the Kane-Mele model in the presence of Rashba interaction. As we have already noted,
$\pm$ and $\uparrow  \downarrow$ label the Dirac points and the $S_z$ eigenvalues.   
They obtained 
\begin{eqnarray*}
N^{\ssSC}_{\pm \uparrow  }=K_{\pm \uparrow }(\infty ) -K_{\pm \uparrow  }(0)&=&\frac{1}{2}, \\
N^{\ssSC}_{\pm \downarrow }=K_{\pm  \downarrow }(\infty ) -K_{\pm  \downarrow }(0)&=&-\frac{1}{2},
\end{eqnarray*}
where 
\begin{equation}
\begin{array}{ccccc}
	K_{+ \uparrow }(p) &= &-K_{+\downarrow }(p ) &= &F_2(p)F_4(p), \\  
K_{- \uparrow }(p)&= &-K_{-\downarrow }(p ) &= &F_6(p)F_8(p). 
\end{array}
\label{on}
\end{equation}
Now,  the ``total spin Chern number" relevant to obtain the spin current can be defined as follows:
\begin{equation}
\label{SC}
N^{\ssSC}= N^{\ssSC}_{+\uparrow  }+N^{\ssSC}_{- \uparrow  }-N^{\ssSC}_{- \downarrow }-N^{\ssSC}_{-\downarrow }=2.
\end{equation}
Although the momentum dependence  of  $R(p)$  and $K(p)$ given in (\ref{infin}) and (\ref{on}) is not the same, the  numerical results  (\ref{br}) and 
(\ref{SC}) suggest that
$$
\sigma_\ssSH\approx \frac{1}{4\pi} N^{\ssSC}.
$$
Obviously, this suggested relation between the coefficient of effective action $C_s,$ and the  ``total spin Chern number" $N^{\ssSC}$ needs 
further clarifications.

We only  dealt with non-interacting electrons. When electron-electron interactions on the honeycomb lattice are introduced in terms of the Hubbard model, the third component of spin is still a good quantum number in the absence of Rashba spin-orbit interactions \cite{rlh,sfr}.  Because of being  a many-body system it  is not evident how to incorporate Hubbard interactions  into the field theory of Kane-Mele model. 
For the mean-field decoupled Hubbard interactions the electron-electron interaction results in shifting the intrinsic spin orbit coupling $\dso$ with a constant related to the Hubbard onsite energy $U,$ 
 due to the properties of the Hubbard model $(\dso=0,\lambda_\ssR=0)$ \cite{rlh}. Hence, when we switch on Rashba spin-orbit interactions, the related field theory for the mean-field decoupled Hubbard interactions will be described by a Lagrangian density  similar to (\ref{lag}) up to some constants.  On the other hand, in \cite{gm} a field theory of the Kane-Mele model with Hubbard interactions  was proposed in terms of some auxiliary fields which are associated with spin gauge field components. Although this formalism reproduces the original model on-shell, how one should  take into account the loop contributions of the auxiliary fields is not clear. 

We demonstrated that response of the quantum spin Hall insulator in the presence of  Rashba interaction can be obtained 
from the effective action  of the external electromagnetic and spin fields  (\ref{Ef}). Therefore the materials analogous to graphene yield   the predictions which are independent of the their detailed structure as far as the underlying Hamiltonian is given by the Kane-Mele model. Their response can be studied within the $BF$-type topological field theory of the external fields. 

\setcounter{equation}{0}
\section{Appendix}
Calculation of the coefficient $C_s$ is straightforward, though it is very cumbersome. Here we would like to present the essential steps of the calculation of $C_s.$ First of all one can observe that
(\ref{c1+c2}) can be expressed in terms of $P^{MN}\equiv P^MS_zP^N+P^NS_zP^M,$ so that  it can be amalgamated with (\ref{c33})
 to write  $C_s$  as
\begin{eqnarray*}
\label{cssum}
C_{s} &=&-\frac{i}{48\pi^2}\int d^2p\tr\Big\{P^{57}\left[(1+\frac{E_1-E_7}{E_1-E_5}+\frac{E_1-E_5}{E_1-E_7}){\ca}^1+(1+\frac{E_3-E_7}{E_3-E_5}+\frac{E_3-E_5}{E_3-E_7}){\ca}^3\right]\\
&&+P^{68}\left[(1+\frac{E_2-E_8}{E_2-E_6}+\frac{E_2-E_6}{E_2-E_8}){\ca}^2+(1+\frac{E_4-E_8}{E_4-E_6}+\frac{E_4-E_6}{E_4-E_8}){\ca}^4\right]\\
&&-P^{13}\left[(1+\frac{E_5-E_3}{E_5-E_1}+\frac{E_5-E_1}{E_5-E_3}){\ca}^5+(1+\frac{E_7-E_3}{E_7-E_1}+\frac{E_7-E_1}{E_7-E_3}){\ca}^7\right]\\
&&-P^{24}\left[(1+\frac{E_6-E_4}{E_6-E_2}+\frac{E_6-E_2}{E_6-E_4}){\ca}^6+(1+\frac{E_8-E_4}{E_8-E_2}+\frac{E_8-E_2}{E_8-E_4}){\ca}^8\right]\\
&&+P^{35}\left[(\frac{E_1-E_3}{E_5-E_3}+\frac{E_1-E_3}{E_1-E_5}){\ca}^1-(\frac{E_7-E_5}{E_3-E_5}+\frac{E_7-E_5}{E_7-E_3}){\ca}^7\right]\\
&&+P^{46}\left[(\frac{E_2-E_4}{E_6-E_4}+\frac{E_2-E_4}{E_2-E_6}){\ca}^2-(\frac{E_8-E_6}{E_4-E_6}+\frac{E_8-E_6}{E_8-E_4}){\ca}^8\right]\\
&&+P^{17}\left[(\frac{E_3-E_1}{E_7-E_1}+\frac{E_3-E_1}{E_3-E_7}){\ca}^3-(\frac{E_5-E_7}{E_1-E_7}+\frac{E_5-E_7}{E_5-E_1}){\ca}^5\right]\\
&&+P^{28}\left[(\frac{E_4-E_2}{E_8-E_2}+\frac{E_4-E_2}{E_4-E_8}){\ca}^4-(\frac{E_6-E_8}{E_2-E_8}+\frac{E_6-E_8}{E_6-E_2}){\ca}^6\right] \Big\} .
\end{eqnarray*} 
Making use of the  polar coordinates (\ref{pol}) and the definitions (\ref{f}) one can show that it can be written in the form
\begin{eqnarray*}
C_{s} &=&-\frac{1}{12\pi^2}\int d^2p\Big[\frac{F_5^2F_7^2pA(5,7)}{(\dso-E_1)(\dso-E_5)(\dso-E_7)}\\
& &(1+\frac{E_1-E_7}{E_1-E_5}+\frac{E_1-E_5}{E_1-E_7})(E_5-E_1) \partial_p F_1^2\\
&&-\frac{F_6^2F_8^2C(6,8)}{p(\dso-E_2)}(1+\frac{E_2-E_8}{E_2-E_6}+\frac{E_2-E_6}{E_2-E_8})(E_6-E_2)\partial_p F_2^2\\
&&+\frac{F_5^2F_7^2pA(5,7)}{(\dso-E_3)(\dso-E_5)(\dso-E_7)}(1+\frac{E_3-E_7}{E_3-E_5}+\frac{E_3-E_5}{E_3-E_7})(E_7-E_3)\partial_p  F_3^2\\
&&-\frac{F_6^2F_8^2C(6,8)}{p(\dso-E_4)}(1+\frac{E_4-E_8}{E_4-E_6}+\frac{E_4-E_6}{E_4-E_8})(E_8-E_4)\partial_p  F_4^2\\
&&+\frac{F_1^2F_3^2pA(1,3)}{(\dso-E_1)(\dso-E_3)(\dso-E_5)}(1-\frac{E_5-E_3}{E_1-E_5}-\frac{E_5-E_1}{E_3-E_5})(E_5-E_1)\partial_p  F_5^2 \\
&&-\frac{F_2^2F_4^2C(2,4)}{p(\dso-E_6)}(1-\frac{E_6-E_4}{E_2-E_6}-\frac{E_6-E_2}{E_4-E_6})(E_6-E_2)\partial_p  F_6^2\\
&&+\frac{F_1^2F_3^2pA(1,3)}{(\dso-E_1)(\dso-E_3)(\dso-E_7)}(1-\frac{E_7-E_3}{E_1-E_7}-\frac{E_7-E_1}{E_3-E_7})(E_7-E_3)\partial_p F_7^2\\
&&-\frac{F_2^2F_4^2C(2,4)}{p(\dso-E_8)}(1-\frac{E_8-E_4}{E_2-E_8}-\frac{E_8-E_2}{E_4-E_8})(E_8-E_4)\partial_p  F_8^2\\
&&+\frac{F_3^2F_5^2pA(3,5)}{(\dso-E_1)(\dso-E_3)(\dso-E_5)}(\frac{E_1-E_3}{E_5-E_3}+\frac{E_1-E_3}{E_1-E_5})(E_5-E_1)\partial_p F_1^2\\
&&-\frac{F_4^2F_6^2C(4,6)}{p(\dso-E_2)}(\frac{E_2-E_4}{E_6-E_4}+\frac{E_2-E_4}{E_2-E_6})(E_6-E_2)\partial_p  F_2^2\\
&&+\frac{F_1^2F_7^2pA(1,7)}{(\dso-E_1)(\dso-E_3)(\dso-E_7)}(\frac{E_3-E_1}{E_7-E_1}+\frac{E_3-E_1}{E_3-E_7})(E_7-E_3)\partial_p  F_3^2\\
&&-\frac{F_2^2F_8^2C(2,8)}{p(\dso-E_4)}(\frac{E_4-E_2}{E_8-E_2}+\frac{E_4-E_2}{E_4-E_8})(E_8-E_4)\partial_p  F_4^2\\
&&+\frac{F_1^2F_7^2pA(1,7)}{(\dso-E_1)(\dso-E_5)(\dso-E_7)}(\frac{E_5-E_7}{E_1-E_7}-\frac{E_5-E_7}{E_1-E_5})(E_5-E_1)\partial_p  F_5^2\\
&&-\frac{F_2^2F_8^2C(2,8)}{p(\dso-E_6)}(\frac{E_6-E_8}{E_2-E_8}-\frac{E_6-E_8}{E_2-E_6})(E_6-E_2)\partial_p  F_6^2\\
&&+\frac{F_3^2F_5^2pA(3,5)}{(\dso-E_3)(\dso-E_5)(\dso-E_7)}(\frac{E_7-E_5}{E_3-E_5}-\frac{E_7-E_5}{E_3-E_7})(E_7-E_3)\partial_p  F_7^2\\
&&-\frac{F_4^2F_6^2C(4,6)}{p(\dso-E_8)}(\frac{E_8-E_6}{E_4-E_6}-\frac{E_8-E_6}{E_4-E_8})(E_8-E_4)\partial_p  F_8^2\Big],\label{css3}
\end{eqnarray*} 
where we defined 
$$
A(m,n)=2\left[1+\frac{p^2}{(\dso-E_m)(\dso-E_n)}\right],\ C(m,n)=2\left[1+\frac{(\dso-E_m)(\dso-E_n)}{p^2}\right].
$$
Expressing $A(m,n)$ and $C(m,n)$ in terms of the normalization factors (\ref{f}), it can  
further be simplified  as 
\begin{eqnarray*}
C_{s} &=&-\frac{1}{6\pi^2}\int d^2p\Big[\frac{(F_5F_7+F_6F_8)F_6F_8}{p(\dso-E_1)}\left(1+\frac{E_1-E_7}{E_1-E_5}+\frac{E_1-E_5}{E_1-E_7}\right)(E_5-E_1)\partial_pF_1^2\\
&&-\frac{(F_5F_7+F_6F_8)F_6F_8}{p(\dso-E_2)}\left(1+\frac{E_2-E_8}{E_2-E_6}+\frac{E_2-E_6}{E_2-E_8}\right)(E_6-E_2)\partial_pF_2^2\\
&&+\frac{(F_5F_7+F_6F_8)F_6F_8}{p(\dso-E_3)}\left(1+\frac{E_3-E_7}{E_3-E_5}+\frac{E_3-E_5}{E_3-E_7}\right)(E_7-E_3)\partial_pF_3^2\\
&&-\frac{(F_5F_7+F_6F_8)F_6F_8}{p(\dso-E_4)}\left(1+\frac{E_4-E_8}{E_4-E_6}+\frac{E_4-E_6}{E_4-E_8}\right)(E_8-E_4)\partial_pF_4^2\\
&&+\frac{(F_1F_3+F_2F_4)F_2F_4}{p(\dso-E_5)}\left(1-\frac{E_5-E_3}{E_1-E_5}-\frac{E_5-E_1}{E_3-E_5}\right)(E_5-E_1)\partial_pF_5^2\\
&&-\frac{(F_1F_3+F_2F_4)F_2F_4}{p(\dso-E_6)}\left(1-\frac{E_6-E_4}{E_2-E_6}-\frac{E_6-E_2}{E_4-E_6}\right)(E_6-E_2)\partial_pF_6^2\\
&&+\frac{(F_1F_3+F_2F_4)F_2F_4}{p(\dso-E_7)}\left(1-\frac{E_7-E_3}{E_1-E_7}-\frac{E_7-E_1}{E_3-E_7}\right)(E_7-E_3)\partial_pF_7^2\\
&&-\frac{(F_1F_3+F_2F_4)F_2F_4}{p(\dso-E_8)}\left(1-\frac{E_8-E_4}{E_2-E_8}-\frac{E_8-E_2}{E_4-E_8}\right)(E_8-E_4)\partial_pF_8^2\\
&&+\frac{(F_4F_6-F_3F_5)F_4F_6}{p(\dso-E_1)}\left(\frac{E_1-E_3}{E_5-E_3}+\frac{E_1-E_3}{E_1-E_5}\right)(E_5-E_1)\partial_pF_1^2\\
&&-\frac{(F_4F_6-F_3F_5)F_4F_6}{p(\dso-E_2)}\left(\frac{E_2-E_4}{E_6-E_4}+\frac{E_2-E_4}{E_2-E_6}\right)(E_6-E_2)\partial_pF_2^2\\
&&+\frac{(F_2F_8-F_1F_7)F_2F_8}{p(\dso-E_3)}\left(\frac{E_3-E_1}{E_7-E_1}+\frac{E_3-E_1}{E_3-E_7}\right)(E_7-E_3)\partial_pF_3^2\\
&&-\frac{(F_2F_8-F_1F_7)F_2F_8}{p(\dso-E_4)}\left(\frac{E_4-E_2}{E_8-E_2}+\frac{E_4-E_2}{E_4-E_8}\right)(E_8-E_4)\partial_pF_4^2\\
&&+\frac{(F_2F_8-F_1F_7)F_2F_8}{p(\dso-E_5)}\left(\frac{E_5-E_7}{E_1-E_7}-\frac{E_5-E_7}{E_1-E_5}\right)(E_5-E_1)\partial_pF_5^2\\
&&-\frac{(F_2F_8-F_1F_7)F_2F_8}{p(\dso-E_6)}\left(\frac{E_6-E_8}{E_2-E_8}-\frac{E_6-E_8}{E_2-E_6}\right)(E_6-E_2)\partial_pF_6^2\\
&&+\frac{(F_4F_6-F_3F_5)F_4F_6}{p(\dso-E_7)}\left(\frac{E_7-E_5}{E_3-E_5}-\frac{E_7-E_5}{E_3-E_7}\right)(E_7-E_3)\partial_pF_7^2\\
&&-\frac{(F_4F_6-F_3F_5)F_4F_6}{p(\dso-E_8)}\left(\frac{E_8-E_6}{E_4-E_6}-\frac{E_8-E_6}{E_4-E_8}\right)(E_8-E_4)\partial_pF_8^2\Big].
\end{eqnarray*} 
Considering the relations (\ref{eig}) and (\ref{f2}) and performing the $\theta$ integral, $C_{s}$ can be written as,
\begin{eqnarray*}
C_{s}&=&-\frac{2}{3\pi}\int dp\Big[\frac{(F_5F_7+F_6F_8)F_6F_8}{\dso-E_1}(1+\frac{E_1-E_7}{E_1-E_5}+\frac{E_1-E_5}{E_1-E_7})(E_5-E_1)\partial_pF_1^2\\
&&+\frac{(F_5F_7+F_6F_8)F_6F_8}{\dso-E_3}(1+\frac{E_3-E_7}{E_3-E_5}+\frac{E_3-E_5}{E_3-E_7})(E_7-E_3)\partial_pF_3^2\\
&&+\frac{(F_1F_3+F_2F_4)F_2F_4}{\dso-E_5}(1-\frac{E_5-E_3}{E_1-E_5}-\frac{E_5-E_1}{E_3-E_5})(E_5-E_1)\partial_pF_5^2\\
&&+\frac{(F_1F_3+F_2F_4)F_2F_4}{\dso-E_7}(1-\frac{E_7-E_3}{E_1-E_7}-\frac{E_7-E_1}{E_3-E_7})(E_7-E_3)\partial_pF_7^2\\
&&+\frac{(F_4F_6-F_3F_5)F_4F_6}{\dso-E_1}(\frac{E_1-E_3}{E_5-E_3}+\frac{E_1-E_3}{E_1-E_5})(E_5-E_1)\partial_pF_1^2\\
&&+\frac{(F_2F_8-F_1F_7)F_2F_8}{\dso-E_3}(\frac{E_3-E_1}{E_7-E_1}+\frac{E_3-E_1}{E_3-E_7})(E_7-E_3)\partial_pF_3^2\\
&&+\frac{(F_2F_8-F_1F_7)F_2F_8}{\dso-E_5}(\frac{E_5-E_7}{E_1-E_7}-\frac{E_5-E_7}{E_1-E_5})(E_5-E_1)\partial_pF_5^2\\
&&+\frac{(F_4F_6-F_3F_5)F_4F_6}{\dso-E_7}(\frac{E_7-E_5}{E_3-E_5}-\frac{E_7-E_5}{E_3-E_7})(E_7-E_3)\partial_pF_7^2\Big].
\end{eqnarray*} 
Finally by making use of relations like
$$
1-\frac{F_3F_5}{F_4F_6}=\frac{E_1-E_3}{E_1-\dso} ,
$$
(\ref{csint}) is accomplished.

\vspace{1cm}
\begin{center}
{\bf Acknowledgment}	
\end{center}

  We are grateful to Tolga Birkandan for his invaluable helps in numerical calculations.
\newpage
\newcommand{\PRL}{Phys. Rev. Lett. }
\newcommand{\PRB}{Phys. Rev. B }
\newcommand{\PRD}{Phys. Rev. D }

\end{document}